\documentclass[10pt,twocolumn,conference]{IEEEtran}
\usepackage{cite}
\usepackage{amsmath,amssymb,amsfonts}

\usepackage{textcomp}
\usepackage{graphicx}
\usepackage{amssymb}
\usepackage{booktabs}
\usepackage{caption}
\usepackage{subcaption}

\usepackage{booktabs}
\usepackage{graphicx}

\usepackage{multicol}
\usepackage[dvipsnames]{xcolor}
\usepackage{minted}

\usepackage{algorithm}
\usepackage{algpseudocode}
\usepackage{multirow,makecell}
\usepackage{units}

\usepackage{colortbl}

\usepackage{tikz}
\usetikzlibrary{positioning, calc, arrows.meta, bending}
\usetikzlibrary{decorations.markings, decorations.pathreplacing}

\usepackage{enumitem}

\usepackage{mathtools}

\algrenewcommand\algorithmicrequire{\textbf{Input:}}
\algrenewcommand\algorithmicensure{\textbf{Output:}}

\definecolor{mylightblue}{HTML}{9FA8DA}
\definecolor{thiswork}{HTML}{61cbf4}

\definecolor{a_blue}{RGB}{97,203,244}
\definecolor{b_orange}{RGB}{233,113,50}
\definecolor{a_green}{RGB}{71, 212, 90}
\definecolor{dark_blue}{RGB}{11,48,65}

\makeatletter % changes the catcode of @ to 11
\newcommand{\linebreakand}{%
  \end{@IEEEauthorhalign}
  \hfill\mbox{}\par
  \mbox{}\hfill\begin{@IEEEauthorhalign}
}
\makeatother % changes the catcode of @ back to 12

\def\BibTeX{{\rm B\kern-.05em{\sc i\kern-.025em b}\kern-.08em
    T\kern-.1667em\lower.7ex\hbox{E}\kern-.125emX}}
\begin{document}

\title{
Parallel Quadratic Selected Inversion in Quantum Transport Simulation
}

\author{
\IEEEauthorblockN{1\textsuperscript{st} Vincent Maillou}
\IEEEauthorblockA{\textit{D-ITET, ETH Zurich} \\
Zurich, Switzerland \\
0000-0003-4861-3298}
\and
\IEEEauthorblockN{2\textsuperscript{nd} Matthias Bollhofer}
\IEEEauthorblockA{\textit{Institute for Numerical Analysis, TUBS} \\
Braunschweig, Germany \\
0000-0002-8093-5812}
\and
\IEEEauthorblockN{3\textsuperscript{rd} Olaf Schenk}
\IEEEauthorblockA{\textit{Institute of Computing, USI} \\
Lugano, Switzerland \\
0000-0001-8636-1023}
\and

\linebreakand

\IEEEauthorblockN{4\textsuperscript{th} Alexandros Nikolaos Ziogas}
\IEEEauthorblockA{\textit{D-ITET, ETH Zurich} \\
Zurich, Switzerland \\
0000-0002-4328-9751}
\and
\IEEEauthorblockN{5\textsuperscript{th} Mathieu Luisier}
\IEEEauthorblockA{\textit{D-ITET, ETH Zurich} \\
Zurich, Switzerland \\
0000-0002-2212-7972}
}

\maketitle

\begin{abstract}
  Driven by Moore's Law, the dimensions of transistors have been pushed down to the nanometer scale.
  Advanced quantum transport (QT) solvers are required to accurately simulate such nano-devices. 
  The non-equilibrium Green's function (NEGF) formalism lends itself optimally to these tasks, but it is computationally very intensive, involving the selected inversion (SI) of matrices and the selected solution of quadratic matrix (SQ) equations. 
  Existing algorithms to tackle these numerical problems are ideally suited to GPU acceleration, e.g., the so-called recursive Green's function (\textsc{RGF}) technique, but they are typically sequential, require block-tridiagonal (BT) matrices as inputs, and their implementation has been so far restricted to shared memory parallelism, thus limiting the achievable device sizes. 
  To address these shortcomings, we introduce distributed methods that build on \textsc{RGF} and enable parallel selected inversion and selected solution of the quadratic matrix equation. 
  We further extend them to handle BT matrices with arrowhead, which allows for the investigation of multi-terminal transistor structures. 
  We evaluate the performance of our approach on a real dataset from the QT simulation of a nano-ribbon transistor and compare it with the sparse direct package \textsc{PARDISO}. When scaling to 16 GPUs, our fused SI and SQ solver is 5.2$\times$ faster than the SI module of \textsc{PARDISO} applied to a device 16$\times$ shorter.
  %Additionally, when compared to the reference \textsc{RGF} algorithm, our approach allows scaling beyond the memory capacity of a single node/GPU. 
  These results highlight the potential of our method to accelerate NEGF-based nano-device simulations.
\end{abstract}

\begin{IEEEkeywords}
\underline{Selected inversion}, Semiconductor device modeling, Distributed algorithms, Linear algebra%, Sparse matrices
\end{IEEEkeywords}

\section{Introduction}~\label{sec:intro}
To keep up with Moore's Law and improve the performance of nanoscale transistors from one generation to the next one, device engineers have come up with innovative solutions, e.g., strain engineering, high-$\kappa$ dielectric layers, or three-dimensional fin field-effect transistors (FinFETs). 
% -------------------------------------
\begin{figure}[t]
    \centering
    \includegraphics[width=\columnwidth]{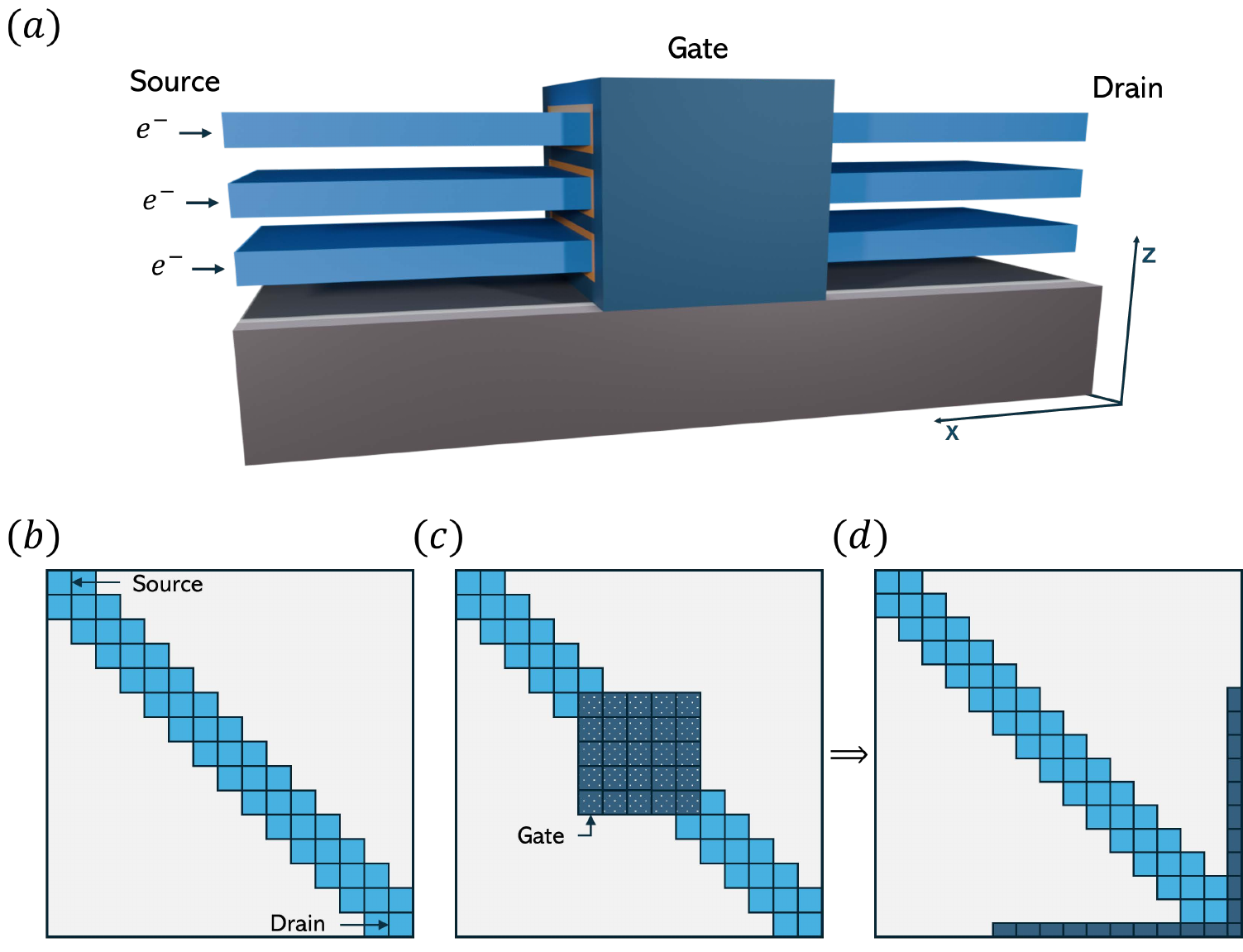}
    \caption{(a) Schematic view of a nano-ribbon field-effect transistor with a gate-all-around configuration~\cite{agrawal_silicon_2024, zhang_gate-all-around_2022, huang_3-d_2020}. Three ribbons are stacked on top of each other, forming the device channel. Electrons (or holes) can be injected into the device structure at the source (left), drain (right), and gate (middle) contacts and then exit at one of these locations. (b)-(d) Typical sparsity patterns of the Hamiltonian matrices that arise in the quantum transport simulation of such devices. (b) Block-tridiagonal matrix describing a classical two-terminal transistor channel (only source and drain). (c) Block-tridiagonal matrix with a large, but sparse central block, further accounting for electron/hole injection and collection at the central gate contact(s). (d) Reordering of the matrix in (c) into a block-tridiagonal with arrowhead sparsity pattern.}
    \label{fig:hamiltonians}
\end{figure}
% -------------------------------------
Crucially, the gate electrodes of transistors should optimally control the flow of charged carriers (electrons or holes) between a source and drain contact, while preventing so-called ``gate leakage currents.'' 
Satisfying these conditions allows to minimize their standby power consumption and achieve high-speed digital switching between logic 0 and 1. 
Silicon FinFETs, the workhorse of the semiconductor industry since 2011, have done so for more than 10 years, but their triple-gate architecture does not provide the necessary level of control at gate lengths of 15 nm and below. 
This is why they are gradually replaced by nano-ribbon (or nano-sheet) field-effect transistors (NR-FET) with a gate-all-around (GAA) configuration~\cite{loubet_stacked_2017, mukesh_review_2022, huang_3-d_2020}, as illustrated in Fig.~\ref{fig:hamiltonians}.
% -------------------------------------
\begin{table*}[t]
\centering
\resizebox{\textwidth}{!}{%
\begin{tikzpicture}
\node (table) {
\begin{tabular}{@{}clcccccccc@{}}
\toprule
\multicolumn{2}{c}{\multirow{2}{*}{\textbf{Solver Libraries}}} &
  \multicolumn{3}{c}{\textbf{Operation}} &
  \multicolumn{2}{c}{\textbf{Sparsity}} &
  \multicolumn{3}{c}{\textbf{Backend}} \\ \cmidrule(l){3-10} 
\multicolumn{2}{c}{} &
  \textbf{Decomposition} &
  \textbf{Selected Inversion} &
  \textbf{Selected Solution} &
  \textbf{BT} &
  \textbf{BTA} &
  \textbf{CPU} &
  \textbf{GPU} &
  \textbf{Distributed Memory} \\ \midrule
\multicolumn{2}{c}{\cellcolor[HTML]{61cbf4} This work} &
  \cellcolor[HTML]{61cbf4} $\checkmark$ &
  \cellcolor[HTML]{61cbf4} $\checkmark$ &
  \cellcolor[HTML]{61cbf4} $\checkmark$ &
  \cellcolor[HTML]{61cbf4} $\checkmark$ &
  \cellcolor[HTML]{61cbf4} $\checkmark$ &
  \cellcolor[HTML]{61cbf4} $\checkmark$ &
  \cellcolor[HTML]{61cbf4} $\checkmark$ &
  \multicolumn{1}{c}{\cellcolor[HTML]{61cbf4} $\checkmark$} \\ \cmidrule(lr){1-10}
\multirow{5}{*}{\rotatebox{90}{\makecell{Libraries}}} &  
  \textsc{Serinv}~\cite{maillou_serinv_2025} &
  PO &
  PO &
   &
  $\checkmark$ &
  $\checkmark$ &
  $\checkmark$ &
  $\checkmark$ &
  $\checkmark$
   \\
  &
   \textsc{PARDISO}$^*$~\cite{schenk_pardiso_2001, schenk_solving_2004} &
  $\checkmark$ &
  $\checkmark$ &
   &
  \multicolumn{2}{c}{General} &
  $\checkmark$ &
   &
   \\
 &
  \textsc{MUMPS}~\cite{amestoy_mumps_2001, amestoy_multifrontal_2000, amestoy_fully_2001} &
  $\checkmark$ &
  $\thicksim$ &
   &
  \multicolumn{2}{c}{General} &
  $\checkmark$ &
   &
  $\checkmark$ \\
 &
  \textsc{ScaLAPACK}~\cite{blackford1997scalapack} &
  $\checkmark$ &
   &
   &
  \multicolumn{2}{c}{Banded / Dense} &
  $\checkmark$ &
   &
  $\checkmark$ \\
 &
  \textsc{cuSolverMp}$^*$~\cite{cusolvermp} &
  $\checkmark$ &
   &
   &
  \multicolumn{2}{c}{Dense} &
   &
  $\checkmark$ &
  $\checkmark$ \\ \cmidrule(lr){1-10} 
\multirow{4}{*}{\rotatebox{90}{\makecell{Alg.\\w. impl}}} &
  \textsc{RGF}~\cite{svizhenko_two-dimensional_2002, petersen_block_2008} &
  $\checkmark$ &
  $\checkmark$ &
  $\checkmark$ &
  $\checkmark$ &
   &
  $\checkmark$ &
  $\checkmark$ &
   \\
 &
  $INLA_{BTA}$~\cite{gaedke-merzhauser_integrated_2023} &
  PO &
  PO &
   &
  $\checkmark$ &
  $\checkmark$ &
   &
  $\checkmark$ &
   \\
 &
  \textsc{SplitSolve}~\cite{calderara_splitsolve_2016} &
  $\checkmark$ &
  $\thicksim$ &
   &
  $\checkmark$ &
   &
   &
  $\checkmark$ &
  $\checkmark$ \\
 &
  \textsc{PSellInv}~\cite{10.1145/1916461.1916464, jacquelin_pselinv_2018} &
  $\checkmark$ &
  $\checkmark$ &
   &
  \multicolumn{2}{c}{General} &
  $\checkmark$ &
   &
  $\checkmark$ \\ \cmidrule(lr){1-10}
\multirow{2}{*}{\rotatebox{90}{\makecell{Alg.\\w.o.\\impl}}} &
  \textsc{FIND}~\cite{li_computing_2008, li_optimization_2009, li_extension_2012, li_fast_2013} &
  $\checkmark$ &
  $\checkmark$ &
  DIAG &
  \multicolumn{2}{c}{General} &
  $\checkmark$ &
   &
  $\checkmark$ \\
 % &
 %  \textsc{PSR}~\cite{petersen_hybrid_2009} &
 %  $\checkmark$ &
 %  $\checkmark$ &
 %   &
 %  $\checkmark$ &
 %   &
 %  $\checkmark$ &
 %   &
 %  $\checkmark$ \\
 &
  \textsc{PSR},\textsc{P-DIV/Spikes}~\cite{petersen_hybrid_2009,polizzi_parallel_2006,cauley_distributed_2011, spellacy_partial_2018} &
  $\checkmark$ &
  $\checkmark$ &
   &
  $\checkmark$ &
   &
  $\checkmark$ &
   &
  $\checkmark$ \\ \bottomrule
\end{tabular}
};
\draw[draw=black, fill=yellow, thick, dashed, rounded corners=3pt, fill opacity=0.3] ($(table.north west) + (11.1, -1.03)$) rectangle ($(table.north west) + (11.7, -3.95)$);
\draw[olive, very thick, dashed, -{Stealth[bend]}] ($(table.north west) + (14.3, -3.6)$) to[out=25, in=-25] ($(table.north west) + (14.3, -1.35)$);
\draw[olive, very thick, dashed, -{Stealth[bend]}] ($(table.north west) + (18.4, -3.6)$) to[out=25, in=-25] ($(table.north west) + (18.4, -1.35)$);
\end{tikzpicture}
}
\footnotesize\\
\textbf{Notes}: $PO$: Positive-definite matrix, DIAG: Limited to diagonal entries, $\thicksim$: Not suitable for this task.
\caption{Summary of the algorithms and/or libraries available for the selected inversion ($AX=I)$ and selected solution of the quadratic equation ($AXA^{\dagger}=B$) problems of BT and BTA structured sparse matrices. The availability of CPU or GPU implementations is indicated, together with the parallelization model.
 Our work is highlighted in blue and novelties over the well-established \textsc{RGF} algorithm are indicated by the dashed arrows. We denote with a $*$ closed-source packages.}
\label{tab:litterature_position}
\end{table*}
% -------------------------------------

Several experimental transistor developments have been supported, and in some cases, even enabled, by physics-based technology computer-aided design (TCAD) tools \cite{wu_2013,kwon_2021,stettler_2021}. 
Such device simulators can reveal the intrinsic mechanisms that affect the functionality of fabricated components, while also providing reliable design guidelines to create new ones. 
To be of practical relevance, modern TCAD packages should work from first principles (no empirical input parameters) \cite{stokbro_2002}, account for electron/hole injection at all contacts \cite{luisier_2008}, offer a quantum mechanical description of the simulation domain, include the most important sources of scattering that limit the performance of nanoscale transistors, in particular surface roughness \cite{wang_2005}, electron-phonon~\cite{frederiksen_2007} and electron-electron~\cite{thygesen_nonequilibrium_2007} interactions, and treat each atom composing the structures of interest individually.

The non-equilibrium Green's function (NEGF) method naturally allows for the combination of all these effects within a single framework~\cite{Datta2000}.
However, due to its high computational burden, it is usually limited to small systems composed of hundreds of atoms, with a few notable exceptions, e.g., in~\cite{10.1145/3295500.3357156}. 
The core operation of NEGF consists of computing different types of Green's functions, $A G^R=I$ for the retarded component $G^R$ and $A G^{<,>} A^{\dagger} = \Sigma^{<,>}$ for its lesser $G^<$ and greater $G^>$ counterparts from which most observables (carrier and current densities) can be derived~\cite{Datta2000}. 
Here, the matrix $A$, of size $N_{AO}\times N_{AO}$ ($N_{AO}$ is the total number of atomic orbitals in the device structure), is directly proportional to the Hamiltonian matrix $H$, which encompasses all material properties. 
The self-energy matrix $\Sigma^{<,>}$, also of size $N_{AO}\times N_{AO}$, accounts for the coupling to semi-infinite contacts (open boundary conditions) and for scattering mechanisms. 

Assuming that electrons (or holes) enter the simulation domain only at the source and drain contacts in Fig.~\ref{fig:hamiltonians}(a) and that the maximum distance between two interacting particles (electron-phonon or electron-electron) does not exceed a cut-off radius $r_{cut}$, $A$ and $\Sigma^{<,>}$ are structured sparse matrices with a block-tridiagonal (BT) shape, as depicted in Fig.~\ref{fig:hamiltonians}(b). 
Additionally, only selected entries of the Green's function matrices $G^{R,<,>}$ are needed to compute the desired observables, i.e., those corresponding to the initial sparsity pattern of $A$ and $\Sigma^{<,>}$~\cite{PhysRevB.111.195421}. 
However, if evaluated through standard methods, the Green's functions become dense matrices during the computation.
For example, the inversion of $A$ to get $G^R$ leads to a full matrix, with the majority of the entries being unnecessary.
As the dimensions of the considered transistors can be large ($N_{AO}\propto O(10^5-10^6)$), the densification occurring during the calculation of $G^{R,<,>}$ can make such operations computationally and memory-wise unfeasible.
Hence, from a numerical point of view, modeling nanoscale devices with NEGF greatly benefits from selected linear algebra methods, such as \emph{selected inversion} and \emph{selected solution}, that directly, and only, produce the desired elements of the Green's function matrices. 
That is why the device modeling community has developed various approaches to compute selected entries of $G^{R,<,>}$, the golden standard being the so-called \emph{recursive Green's function} (\textsc{RGF}) algorithm~\cite{lake_1997,svizhenko_two-dimensional_2002}. 
Notably, \textsc{RGF} allows for the conjoint selected solution of the quadratic equation $A G^{<,>} A^{\dagger} = \Sigma^{<,>}$ for the lesser and greater Green's functions, together with the selected inversion of $A$ for the retarded Green's function.
These features are further discussed in Section~\ref{sec:background}.

\textsc{RGF} readily applies to transistor structures where the blocks in the matrix $A$ have different sizes. 
This is the case, for instance, when the device's cross-section varies between source and drain or when electrons have a non-negligible probability of tunneling through the insulator layer separating the transistor channel from the gate contacts. The induced gate leakage currents should be minimized to avoid large stand-by power consumption. To account for them, electrons must be injected and collected at the gate electrodes, which introduces a large, highly sparse block in the center of the matrix $A$, as illustrated in Fig.~\ref{fig:hamiltonians}(c). 
This block results from the open boundary conditions that connect all atoms attached to the gate regions with the central device region~\cite{luisier_2008}. 
Since \textsc{RGF} necessitates the inversion of all diagonal blocks of $A$, the presence of a large unit in the middle of this matrix leads to a sub-optimal computational performance and high memory consumption. 
With appropriate permutation, this matrix can nevertheless be transformed into a block-tridiagonal with arrowhead (BTA) structure (Fig.~\ref{fig:hamiltonians}(d)), which possesses an advantageous computational pattern.

While both BT and BTA matrices play an important role in quantum transport simulations, the \textsc{RGF} algorithm only works for BT sparsity patterns, preventing the investigation of realistic atomic systems with more than two contacts.
If implemented on modern GPUs it can return selected entries of $G^{R,<,>}$ for homogeneous devices made of up to $N_{A0}\simeq$40,000 orbitals~\cite{sc24}, depending on the ratio between the structure's cross-section ($y$-$z$ plane in Fig.~\ref{fig:hamiltonians}(a)) and its length ($x$ axis). 
Still, due to its block-sequential nature, \textsc{RGF} is limited to shared memory (SM) parallelism at the block level, hindering the simulation of atomic systems whose $A$ matrix does not fit into the memory of a single node.

\subsection{Related work}
Several attempts have been made to parallelize the calculation of retarded and lesser/greater Green's functions.
Distributed memory (DM) algorithms for the parallel solution of the retarded Green's function $G^R$ include \textsc{PSellInv}~\cite{10.1145/1916461.1916464}, \textsc{PSR}~\cite{petersen_hybrid_2009}, and \textsc{SplitSolve}~\cite{calderara_splitsolve_2016}.
While efficient, these methods are limited to ballistic transport simulations as they do not return $G^<$ and $G^>$. These quantities are necessary to describe scattering mechanisms~\cite{Datta2000}.
Efforts to parallelize the solution of the quadratic matrix equation, which returns $G^{<,>}$, have also been undertaken, but the proposed approaches assume that the scattering self-energy matrices $\Sigma^{<,>}$ exhibit a (block) diagonal sparsity pattern. 
Notable examples include \textsc{FIND}~\cite{li_fast_2013} and P-DIV~\cite{cauley_distributed_2011}.
Moreover, none of these methods have been ported to GPU, limiting their usefulness on today's hybrid supercomputer.
Alternatively, state-of-the-art sparse direct solvers such as \textsc{PARDISO}~\cite{schenk_pardiso_2001} or \textsc{MUMPS}~\cite{amestoy_mumps_2001} can be leveraged to handle the large linear systems encountered in QT.
However, they both face significant limitations when it comes to selected solutions of quadratic matrix equations; they do not support GPU acceleration, and \textsc{PARDISO} lacks DM capabilities (see Table~\ref{tab:litterature_position}). 
On the other hand, distributed structured solvers like the block-banded diagonally dominant routine (PZGBSV) of \textsc{ScaLAPACK} may not offer selected inversion capabilities, but they generally scale relatively well~\cite{blackford1997scalapack}. 
Nevertheless, owing to their very disadvantageous computational complexity and high memory constraints, they are ill-suited to perform QT simulations. 
These issues are further detailed in Section~\ref{sec:other_approaches}, and a comprehensive overview of the presented algorithms is provided in Table~\ref{tab:litterature_position}.

\subsection{Our contribution}
In this work, we first present an extension of the \textsc{RGF} algorithm to BTA sparse matrices.
We then introduce novel parallel algorithms derived from \textsc{RGF} to perform the selected inversion and compute the selected solution of the quadratic matrix equation in distributed-memory environments.
The proposed developments take advantage of block Gaussian elimination and domain decomposition techniques, enabling NEGF-based quantum transport simulations of large atomic structures, including scattering effects and multi-terminal injection capabilities.
Our solver is adapted to both CPU and GPU architectures, with its main features highlighted in blue in Table~\ref{tab:litterature_position}.
Our approach is essential to scale nano-devices simulation beyond the \textsc{RGF} memory constraint of a single node/GPUs.
The main contributions of this work are the following:
\begin{itemize}
    \item Extension of the \textsc{RGF} algorithm to BTA sparse matrices;
    \item Derivation of distributed memory algorithms for selected solution of the quadratic matrix equation in case of BT and BTA sparse matrices;
    \item Porting of the derived methods to GPUs;
    \item Performance evaluation (weak scaling) of our implementations on a NR-FET transistor dataset, with comparison to \textsc{PARDISO}.
    \item Evaluation of the parallel efficiency (weak and strong scaling) of our GPU accelerated implementation on up to 32 GPUs.
\end{itemize}

The paper is organized around five sections. 
We introduce in Section~\ref{sec:background} the required mathematical and algorithmic background, and present the \textsc{RGF} algorithm.
Our novel methods are derived in Section~\ref{sec:methods} before conducting numerical evaluation of the implemented sequential and distributed codes in Section~\ref{sec:evaluation}. 
Finally, conclusions are drawn in Section~\ref{sec:conclusion}.

\section{Background}\label{sec:background}
\subsection{Problem Formulation}~\label{ssec:problem_formulation}
Selected linear algebra refers to the class of numerical methods that do not only account for the sparsity of the input data (e.g., the system matrix $A$ in $AX=B$), but also of the solution space (e.g., the output matrix $X$). 
The key benefit of such methods is that they only return the desired elements of $X$ and avoid a dense expression of the solution, which could make its computation and/or storage unfeasible.
In case of quantum transport simulations within the NEGF framework, two equations are of particular interest: $AX=I$ and $AXA^\dagger=B$. 
In both cases, $A$ and $B$ are sparse matrices that typically exhibit a BT or BTA sparsity pattern. 
Commonly, only the entries of $X$ matching the non-zero sparsity patterns of $A$ and $B$ (both matrices have the same structure) are needed to extract the desired information.
The BT (BTA) pattern of $A$, $B$, and $X$ can be ideally tiled into a block-tridiagonal (with arrowhead) pattern.
This tiling assumes $n$ square diagonal blocks. Here, for simplicity, we only consider the case where all of them are uniform and of size $b$. This parameterization is shown in Fig.~\ref{fig:schur_bt} (a).
In case of BTA, the size of the arrowhead is $a$ (see Fig.~\ref{fig:bta_i}). 
These and all other symbols used in this work are defined in Table~\ref{tab:notation}.

The algorithms described in the following sections typically consist of two phases, a matrix decomposition, referred to as \emph{forward-pass}, followed by a combination of triangular solve routines, referred to as \emph{backward-pass}. 
In the general case, numerical stability is ensured if the system matrix exhibits (block-)diagonal dominance (DD) properties~\cite{feingold1962block}, which are fulfilled in quantum transport problems, and more generally in matrices describing nearest-neighbor type of interactions.

\begin{table}[t]
\scriptsize
\centering
\rowcolors{2}{gray!15}{white} % Alternating row colors start from the second row
\begin{tabular}{p{1.78cm}p{6.25cm}}
\toprule
\textbf{Name} & \textbf{Description} \\
\midrule
% First section: Matrix Parameters
\rowcolor{white} \multicolumn{2}{c}{\textbf{Matrix Parameters}} \\ 
\cmidrule(lr){1-2} % Midrule spanning only the description column
$n$ & Number of square blocks forming the main diagonal of a BT(A) matrix (excluding the arrow tip block). \\
$b$ & Size of the square diagonal and upper-/lower-diagonal blocks of a BT(A) matrix (excluding the arrowhead blocks).\\
$a$ & Size of the arrow tip block in a BTA matrix. \\
$N$ & Total size of the BT (BTA) matrix: $nb$ ($nb + a$).\\
$P$ & Number of parallel matrix partitions. When prepended to a name, $P$ indicates a parallel version of the algorithm.\\
% Second section: Algorithm Concepts
\cmidrule(lr){1-2}
\rowcolor{white} \multicolumn{2}{c}{\textbf{Algorithmic Concepts}} \\ 
\cmidrule(lr){1-2} % Midrule spanning only the description column
Block-sequential & Refers to algorithms and routines expressed in matrix-block operations with \emph{sequential} dependencies among them.\\
Forward/Backward Pass & Block-sequential for-loop operating from the top-left or bottom-right blocks of (the partition of) a matrix towards the bottom-right or top-left ones.\\
Fused Operations & Refer to an algorithm that produces conjointly several results from the same set of operands, effectively fusing their operations and reducing the total complexity.\\
\textbf{$/$} (\textbf{$\backslash$}) & Forward (backward) triangular solve\\
\cmidrule(lr){1-2}
\rowcolor{white} \multicolumn{2}{c}{\textbf{Matrix Types and Algorithms}} \\ 
\cmidrule(lr){1-2} % Midrule spanning only the description column
DD & General diagonally dominant (or block-diagonally dominant) matrix.\\
BT (BTA) & Block-tridiagonal (with arrowhead) matrix.\\
SI & Selected Inversion of the system $AX=I$\\
SQ & Selected solution of the Quadratic matrix equation $AXA^{H}=B$.\\
SC & Schur complement algorithm.\\
SCI (SCQ) & Selected inversion (quadratic matrix equation solution) based on Schur complement calculation.\\
\textsc{RGF} & Recursive Green's Function algorithm.\\
\bottomrule
\end{tabular}
\caption{List of the symbols and terms used in this work.}
\label{tab:notation}
\end{table}

\subsection{The \textsc{RGF} Algorithm}~\label{ssec:rgf}
The Recursive Green's Function (\textsc{RGF}) algorithm performs selected inversion and solution of the quadratic matrix equation of systems based on BT matrices. 
It is the \textit{de facto} standard approach for quantum transport calculations within the NEGF formalism, as it allows to compute only selected elements of $X=A^{-1}$ and $X=A^{-1}BA^{-\dagger}$. 

\begin{figure*}[t]
    \centering
    \includegraphics[width=0.9\textwidth]{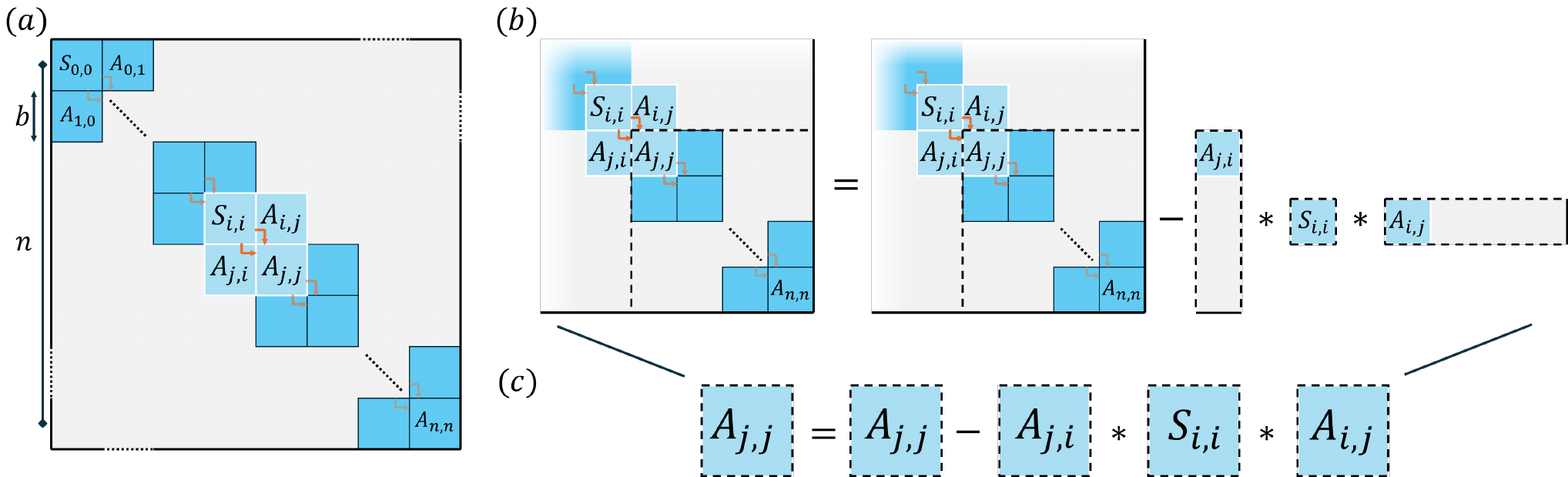}
    \footnotesize\\
    \vspace{0.5em}
    \textbf{Notes}: $j=i+1$.
    \caption{High-level overview of the derivation of the forward pass of the \textsc{RGF} algorithm for a block-tridiagonal (BT) structured sparse matrix. (a) BT sparsity pattern of the input $A$ matrix and blocks involved in the $i$-th step of the forward pass. (b) Slicing of the matrix at the $i$-th step and zoom-in on the block operations involved in the forward update. (c) Reduction of the block operation in (b) by avoiding \textit{a priori} zero computations.}
    \label{fig:schur_bt}
\end{figure*}

From a numerical point of view, \textsc{RGF} relies on a sparse Schur complement $S_A$ (forward pass) of the system matrix and a selected inversion (backward pass), equivalent to a restriction of the algorithm derived by Takahashi~\cite{takahashi_sellinv} to the BT sparsity pattern of $A$. 
To do so, a block decomposition $A=LDU$ is computed in the forward pass, where $L,U$ have unit block diagonal part. For reasons of efficiency, only the block diagonal matrix $D$ is stored whereas the entries of $L$ and $U$ are temporarily computed during the backward pass. Inverting $D$ leads to $S_A=D^{-1}$.
\textsc{RGF} can be extended with a second Schur complement $S_B$ which refers to the diagonal blocks of $S_AL^{-1}BL^{-\dagger}S_A^\dagger$, fusing the solution of the quadratic matrix equation and the selected inversion, thus reducing the overall computational complexity by only returning selected entries of the solution space.
The \textsc{RGF} algorithm is labeled as \emph{recursive} as it iterates forward (backward) during the Schur complement (selected inversion+quadratic matrix equation) over the diagonal blocks of the BT matrix.
At step $i$, a Schur complement is computed and propagated as an update to the next diagonal block. 
This intermediate step is represented for a BT matrix in Fig.~\ref{fig:schur_bt} (a). 
The block operations and the complete description of the \textsc{RGF} forward-pass are described in Alg.~\ref{alg:rgf_schur}. 
The operations related to the selected inversion of $A$ are highlighted in \textcolor{PineGreen}{green}, while those referring to the selected solution of the quadratic matrix equation are marked in \textcolor{RedViolet}{violet}.

\begin{algorithm}[t]
\small
\caption{\textsc{RGF} Schur complement (forward pass)}
\label{alg:rgf_schur}
\begin{algorithmic}[1]
    \Require Block-tridiagonal matrices: $A$ and $B$.
    \For{$n_i = 0$ to $n-2$}
        \State \textcolor{PineGreen}{$S_{A_{i,i}} \gets A_{i,i}^{-1}$}
        \State \textcolor{RedViolet}{$S_{B_{i,i}} \gets S_{A_{i,i}}  B_{i,i}  S_{A_{i,i}}^{\dagger}$}
        \State \textcolor{PineGreen}{$\text{T}^{A}_{1} \gets A_{i+1,i}  S_{A_{i,i}}$}
        \State \textcolor{PineGreen}{$A_{i+1,i+1} \gets A_{i+1,i+1} - \text{T}^{A}_{1}  A_{i,i+1}$}
        \State \textcolor{RedViolet}{$B_{i+1,i+1} \gets B_{i+1,i+1} + A_{i+1,i}  S_{B_{i,i}}  A_{i+1,i}^{\dagger} - B_{i+1,i}  {\text{T}^{A}_{1}}^{\dagger} - \text{T}^{A}_{1}  B_{i,i+1}$}
    \EndFor
    \State \textcolor{PineGreen}{$S_{A_{n-1,n-1}} \gets A_{n-1,n-1}^{-1}$}
    \State \textcolor{RedViolet}{$S_{B_{n-1,n-1}} \gets S_{A_{n-1,n-1}}  B_{n-1,n-1}  S_{A_{n-1,n-1}}^{\dagger}$}
\end{algorithmic}
\begin{tikzpicture}
    \fill[PineGreen] (0,0) rectangle (0.25cm,0.25cm);
    \node[right] at (0.25cm,0.125cm) {Selected Inversion};
    \fill[RedViolet] (3cm,0) rectangle (3.25cm,0.25cm);
    \node[right] at (3.25cm,0.125cm) {Selected Solution Quadratic Eq.};
\end{tikzpicture}
\end{algorithm}
% ---------------------------------------------------------
\begin{algorithm}[t]
\small
\caption{\textsc{RGF} SI and SQ (backward pass)}
\label{alg:rgf_sellinv}
\begin{algorithmic}[1]
    \Require Schur complement matrices: $S_{A}$ and $S_{B}$.
    \State \textcolor{PineGreen}{$X_{A_{n-1,n-1}}\gets S_{A_{n-1,n-1}}$}
    \State \textcolor{RedViolet}{$X_{B_{n-1,n-1}}\gets S_{B_{n-1,n-1}}$}
    \For{$i = n-2$ to $0$}
        % \State j = i + 1
        \State \textcolor{PineGreen}{$\text{T}^{A}_{1} \gets S_{A_{i,i}} A_{i,i+1}$}, \textcolor{PineGreen}{$\text{T}^{A}_{2} \gets X_{A_{i+1,i+1}} A_{i+1,i}$}, %\textcolor{PineGreen}{$\text{T}^{A}_{3} \gets A_{i+1,i}$}
          \State \textcolor{PineGreen}{$X_{A_{i+1,i}} \gets -\text{T}^{A}_{2} S_{A_{i,i}}$}
        \State \textcolor{PineGreen}{$X_{A_{i,i+1}} \gets -\text{T}^{A}_{1} X_{A_{i+1,i+1}}$}
        \State \textcolor{PineGreen}{$X_{A_{i,i}} \gets S_{A_{i,i}} -  \text{T}^{A}_{1} X_{A_{i+1,i}}$}
      \State \textcolor{RedViolet}{$\text{T}^{B}_{1} \gets X_{B_{i+1,i+1}} {\text{T}^{A}_{1}}^{\dagger}$}, \textcolor{RedViolet}{$\text{T}^{B}_{2} \gets S_{B_{i,i}} {\text{T}^{A}_{2}}^{\dagger}$}, \textcolor{RedViolet}{$\text{T}^{B}_{3} \gets \text{T}^{A}_{2} S_{B_{i,i}}$}
        \State \textcolor{RedViolet}{$\text{T}^{B}_{4} \gets S_{A_{i,i}} {B_{i,i+1}} X_{A_{i+1,i+1}}^{\dagger}$}, \textcolor{RedViolet}{$\text{T}^{B}_{5} \gets X_{A_{i+1,i+1}} {B_{i+1,i}} S_{A_{i,i}}^{\dagger}$}
        \State \textcolor{RedViolet}{$X_{B_{i,i+1}} \gets -\text{T}^{A}_{1} X_{B_{i+1,i+1}} - \text{T}^{B}_{2} + \text{T}^{B}_{4}$}
        \State \textcolor{RedViolet}{$X_{B_{i+1,i}} \gets -\text{T}^{B}_{1} - \text{T}^{B}_{3} + \text{T}^{B}_{5}$}
        \State \textcolor{RedViolet}{$X_{B_{i,i}} \gets S_{B_{i,i}} + \text{T}^{A}_{1} \text{T}^{B}_{1} + \text{T}^{A}_{1} \text{T}^{B}_{3} + \text{T}^{B}_{2} {\text{T}^{A}_{1}}^{\dagger} - \text{T}^{A}_{1} \text{T}^{B}_{5} - \text{T}^{B}_{4} {\text{T}^{A}_{1}}^{\dagger}$}
    \EndFor
\end{algorithmic}
\begin{tikzpicture}
    \fill[PineGreen] (0,0) rectangle (0.25cm,0.25cm);
    \node[right] at (0.25cm,0.125cm) {Selected Inversion};
    \fill[RedViolet] (3cm,0) rectangle (3.25cm,0.25cm);
    \node[right] at (3.25cm,0.125cm) {Selected Solution Quadratic Eq.};
\end{tikzpicture}
\end{algorithm}

During the forward-pass, \textsc{RGF} performs at each step an LU decomposition (\emph{getrf}) of the diagonal block $A_{i,i}$, a triangular matrix solve (\emph{trsm}) on the lower (upper) diagonal block $A_{i+1,i}$ ($A_{i,i+1}$), and several general matrix multiplications (\emph{gemm}). 
Additional triangular solves and matrix multiplications are necessary to account for the complement $S_B$ of the quadratic matrix equation. 
When \textsc{RGF} is implemented on GPUs, it may be preferable to compute the inverse of the diagonal blocks and create the forward updates through matrix multiplication instead of triangular solve routines.
This ``trick'' enables pivoting at the block level without the need for panel permutation~\cite{https://doi.org/10.1002/cpe.1829}, and exposes more matrix multiplications, which is crucial for high-performance GPU implementations. 
This formulation is adopted in Alg.~\ref{alg:rgf_schur}.

The backward selected inversion and solution of the quadratic matrix equation are stepping processes from the last diagonal block (lower right) toward the first diagonal block (upper left), reversing the dependencies of the forward pass.
The off-diagonal blocks of the inverse (quadratic equation) $X_{A/B_{i,j}}$ and $X_{A/B_{j,i}}$ can be explicitly computed (Alg.~\ref{alg:rgf_sellinv} lines 6, 7, 9, and 10) or, when only the diagonal blocks are desired, discarded in favor of reduced transient results.

The \textsc{RGF} algorithm faces two main limitations, which hinder the scalability and number of atoms achievable in quantum transport simulations. 
Firstly, it is restricted to BT sparsity patterns. 
In most cases, a dense BT tiling can be adapted to the system matrix describing the device at the expense of extra fill-in and redundant computations with zeroes. 
Given a suited permutation, these structures can often be permuted into a BTA sparsity pattern, reducing the fill-in induced during the forward pass.
Secondly, the \textsc{RGF} algorithm exhibits block sequential dependencies in its forward and backward passes, which restrict its implementations to shared-memory (fine-grained) parallelism, and thus the size of the nano-devices that can be investigated to the memory capacity of a single GPU.

\subsection{Other approaches to solve $AXA^\dagger=B$}\label{sec:other_approaches}
As already mentioned in Section~\ref{sec:intro}, several alternatives to \textsc{RGF} exist to perform selected inversions of a matrix $A$, with dedicated functionalities available in \textsc{PARDISO} and \textsc{MUMPS}. 
The situation is different for the quadratic equation $AXA^\dagger=B$. 
A greedy approach to solve it would be to first do an LU decomposition of the system matrix $A$ and then perform a series of triangular matrix solves on the right-hand-side matrix $B$.
We consider two variants of this method, the first one, presented in Section~\ref{sec:sparse_solver}, relies on a sparse approach for the decomposition of $A$ and a batching of the forward/backward substitution. 
In the second one in Section~\ref{sec:dense_solver}, a dense factorization of $A$ is followed by succinct forward/backward substitutions, which leads to the complete solution $X=A^{-1}BA^{-\dagger}$.

\subsubsection{Sparse Solvers}\label{sec:sparse_solver}
General sparse solvers, including those for banded matrices, maintain sparsity during the decomposition procedure, with potentially extra fill-in.
For banded or block-tridiagonal (BT) matrices, this fill-in is zero.
In case of the selected solution of $AX=I$ and $AXA^{\dagger}=B$, assuming that entries spanning all columns of $X$ are needed, obtaining $X$ based on the $LU$ factorization of $A$ leads to a dense result.
To mitigate this issue, a batched approach can be used, as described in Alg.~\ref{alg:batched_solve}.
After each triangular solve for a given slice $b_i = B[:, i]$, the undesired entries can be masked out, and the operation repeated until all columns of $B$ have been considered.
The computational complexity of this approach is equivalent to a dense inverse of $A$ followed by matrix multiplications $A^{-1}BA^{-\dagger}$. Batching ensures, however, that the total memory required is constrained to the desired solution space.

Additionally, state-of-the-art general sparse solvers suffer from a lack of GPU acceleration.
Being typically based on elimination graphs arising from fill-in reordering methods~\cite{metis_manual, metis_fillin}, they often rely on dense but small tiles~\cite{supernode} (e.g., \emph{supernodes} in the case of \textsc{PARDISO}, \emph{fronts} in the case of \textsc{MUMPS}) where level-3 BLAS kernels can be employed.
The dimension of these tiles makes such solvers suited for multi-threaded CPU execution, but limits the exposed parallelism and thus the performance on GPUs.

\begin{algorithm}[]
\scriptsize
\caption{Sparse/Structured Batched Solve}
\label{alg:batched_solve}
\begin{algorithmic}[1]
    \State $L, U \gets \text{getrf}(A)$
    \For{$i = 0$ to $N$}
        \State $b_i = B[:, i]$
        \State $y_i = U \backslash (L / b_i)$
        \State $x_i = (y_i \backslash L^{-\dagger}) / U^{-\dagger}$
    \EndFor
\end{algorithmic}
\end{algorithm}

\subsubsection{Dense Solvers}\label{sec:dense_solver}
Dense methods involve a dense (or banded) decomposition of the system matrix, and a dense expression of the inverse of $A$.
The produced $A^{-1}$ can then be used to solve the quadratic matrix equation through matrix multiplication with the right-hand-side matrix $B$, as described in Alg.~\ref{alg:batched_solve}.
The result can be sparsified by removing the non-desired entries of the obtained dense solution.
In this approach, the computational complexity amounts to $O(N^3)$, and the memory footprint to $O(N^2)$.

\begin{algorithm}[]
\scriptsize
\caption{Dense Solve}
\label{alg:dense_solve}
\begin{algorithmic}[1]
    \State $L, U \gets \text{getrf}(A)$
    \State $Y = U\backslash(L / B)$
    \State $X = (Y \backslash L^{-\dagger}) / U^{-\dagger} $
\end{algorithmic}
\end{algorithm}

\subsubsection{Summary}\label{sec:summary}
The aforementioned limitations of \textsc{RGF} alternatives to solve $AXA^{\dagger}=B$ make them impractical for real-world use cases.
Their complexity is summarized in Table~\ref{tab:complexity} and compared to the \textsc{RGF} algorithm. 
For completeness, we also present the characteristics of their distributed-memory variants. 
If $b \simeq N$, both the sparse and dense approaches become computationally attractive. 
However, in typical nano-device simulations, we deal with $2b \ll N$ matrices, i.e., their bandwidth is much smaller than their size. 
To handle such cases, we present in this paper new methods that achieve high distributed-memory scalability for both the forward and backward passes of \textsc{RGF}-like algorithms, including the selected solution of the quadratic matrix equation, on CPUs and GPUs.

\begin{table}[h]
\centering
\resizebox{\columnwidth}{!}{%
\begin{tabular}{@{}clcccc@{}}
\toprule
\multicolumn{2}{c}{\multirow{2}{*}{\textbf{Approaches}}} &
  \multicolumn{2}{c}{\begin{tabular}[c]{@{}c@{}}\textbf{Decomposition}\\ \textit{(forward pass)}\end{tabular}} &
  \multicolumn{2}{c}{\begin{tabular}[c]{@{}c@{}}\textbf{Selected Solution Quadratic Eq.}\\ \textit{(backward pass)}\end{tabular}} \\
  \cmidrule(l){3-6}
\multicolumn{2}{c}{}                    & Compute   & Memory    & Compute   & Memory    \\
\midrule
\multirow{3}{*}{\rotatebox{90}{\makecell{Seq.}}} & \textsc{RGF} & $O(nb^3)$ & $O(nb^2)$ & $O(nb^3)$ & $O(nb^2)$ \\
                             & Sparse & $O(nb^3)$ & $O(nb^2)$ & \textcolor{BrickRed}{$O(N^3)$} & $O(nb^2)$ \\
                             & Dense  & \textcolor{BrickRed}{$O(N^3)$} & \textcolor{BrickRed}{$O(N^2)$} & \textcolor{BrickRed}{$O(N^3)$} & \textcolor{BrickRed}{$O(N^2)$} \\
\midrule
\multirow{3}{*}{\rotatebox{90}{\makecell{Dist.}}} & This work & $O(\frac{nb^3}{P}+Pb^3)$ & $O(\frac{nb^2}{P}+Pb^2)$ & $O(\frac{nb^3}{P}+Pb^3)$ & $O(\frac{nb^2}{P}+Pb^2)$ \\
                             & Sparse   & $O(\frac{nb^3}{P}+Pb^3)$ & $O(\frac{nb^2}{P}+Pb^2)$ & $O(N^3/P)$ & $O(\frac{nb^2}{P})$ \\
                             & Dense    & $O(N^3/P)$ & $O(N^2/P)$ & $O(N^3/P)$ & $O(N^2/P)$ \\
\bottomrule
\end{tabular}
}
\caption{High-level summary of the complexity and memory footprint of dense, sparse, and \textsc{RGF} approaches to the selected solution of $AXA^{\dagger}=B$ for a block-tridiagonal matrix with parameters $[n,b]$. Highlighted in \textcolor{BrickRed}{red} are the complexities leading to infeasibility.}
\label{tab:complexity}
\end{table}

\section{Method}\label{sec:methods}
We introduce two methods to compute the selected solution of the quadratic matrix equation $AXA^\dagger=B$. 
First, we extend the \textsc{RGF} algorithm to BTA sparsity patterns, enabling the modeling of a wider class of nanoscale devices.
Secondly, we propose distributed methods based on domain decomposition and reduced system assembly targeting both BT and BTA sparsity patterns.

% ----------------------------------
\begin{figure}[t]
    \centering
    \includegraphics[width=0.98\columnwidth]{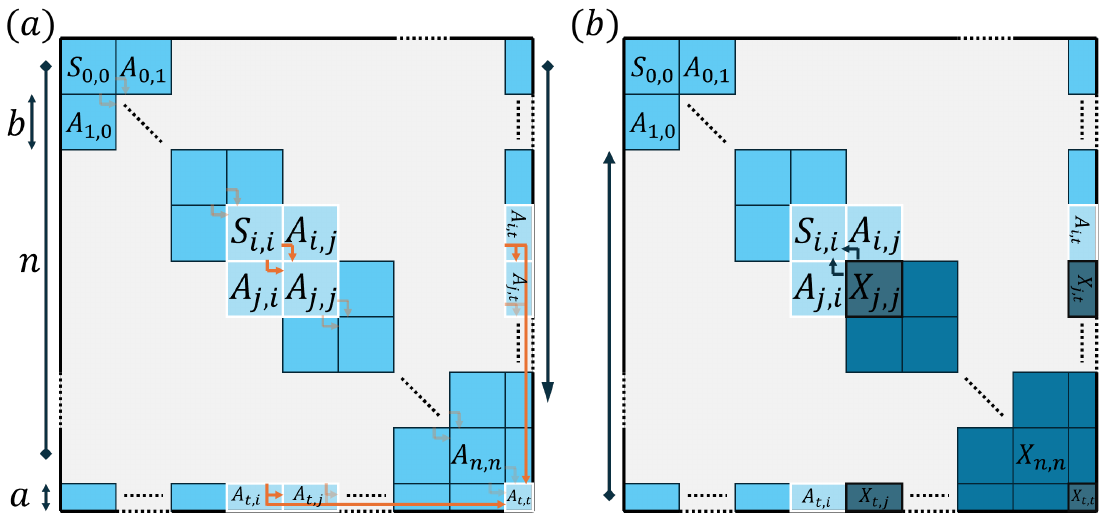}
    \footnotesize\\
    \textbf{Notes}: $j=i+1$.
    \caption{Extension of the \textsc{RGF} algorithm toward matrices with a block-tridiagonal with arrowhead sparsity pattern. (a) Slicing of the system matrix at the $i-th$ step of the forward Schur complement operation. (b) Slicing of the system matrix at the $i-th$ step of the backward selected inversion operation. The \textcolor{b_orange}{orange} and \textcolor{dark_blue}{marine blue} arrows showcase the stepping direction and indicate the block operations performed.}
    \label{fig:bta_i}
\end{figure}
% ----------------------------------

\subsection{Extension of \textsc{RGF} to BTA Matrices}~\label{ssec:bta_schur}
\begin{figure}[b]
    \centering
    \includegraphics[width=0.95\columnwidth]{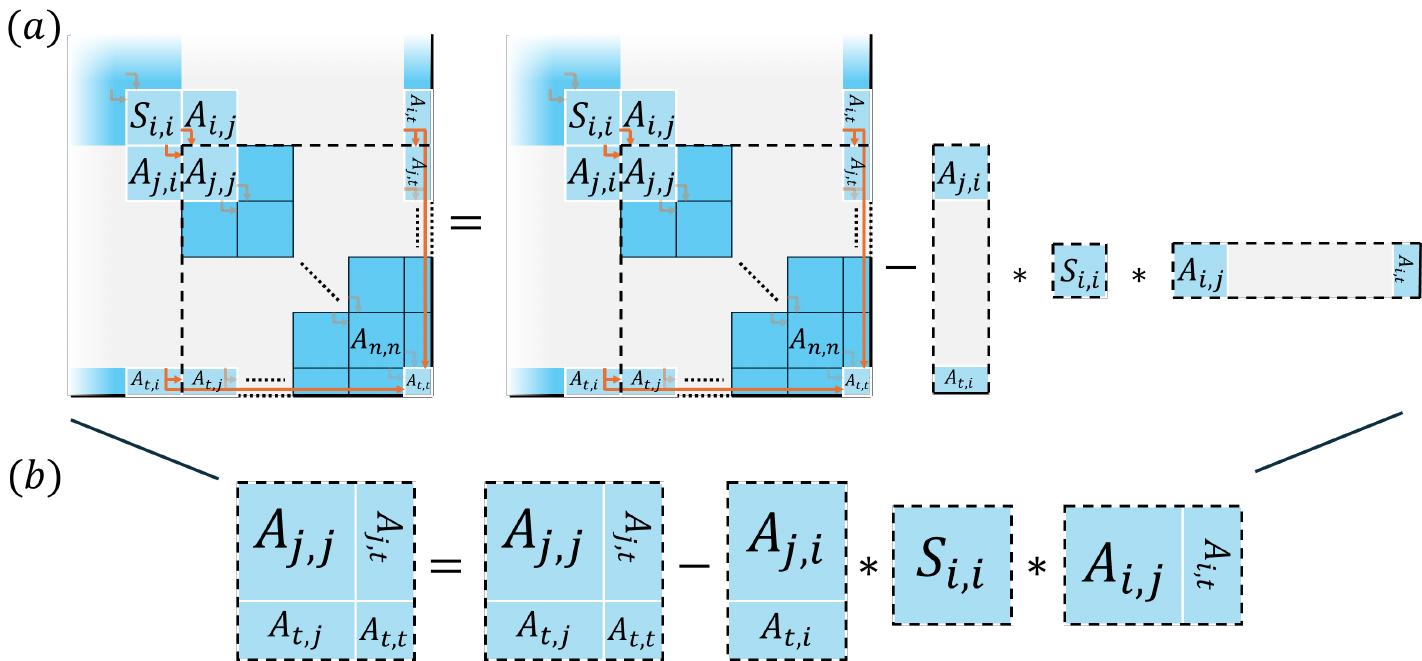}
    \footnotesize\\
    \textbf{Notes}: $j=i+1$.
    \caption{Block operations in the forward update of the extended \textsc{RGF} algorithm ($i$-th step). (a) With all zero operations. (b) After compressing, avoiding \textit{a priori} zero operations.}
    \label{fig:schur_bta_i}
\end{figure}
The \textsc{RGF} algorithm is tailored to matrices with a BT sparsity. 
To treat BTA sparsity patterns, we present a high-level derivation of our approach analogous to the one introduced in Section~\ref{ssec:rgf}, focusing more specifically on the forward Schur complement phase.
This derivation is based on a visual representation of the block-computation arising at a given step $i$.
We consider the slice $A_{i:,i:}$ of the BTA matrix $A$, as presented in Fig.~\ref{fig:bta_i}(a).
The orange arrows indicate the iteration direction, starting from the upper-left diagonal block down to the arrowhead tip.
The light blue overlay highlights the blocks involved during the SC formation and forward update of $A_{j:,j:}$ at step $i$.
% The slicing presented here hold for the forward pass (Schur complement).
The selected inversion/solution of the quadratic matrix equation is based on a mirrored slicing, starting from the last block of the matrix (tip of the arrowhead $A_{t,t,}$) and running backward toward $A_{0, 0}$ (Fig.~\ref{fig:bta_i}(b)).
The marine blue blocks represent the exact entries of the inverse/solution, while the dark overlay highlights the blocks involved during the backward selected inversion $X_{i:,i:}$ at step $j$.

The block derivation of the extension of the \textsc{RGF} algorithm to BTA matrices is presented in Fig.~\ref{fig:schur_bta_i}. 
We devise in sub-plot(a) the level-3 BLAS operations that are performed for each diagonal block at step $i$.
In case of BTA sparsity pattern, these operations only give rise to a fixed number of non-zero block operations, namely the arrowhead blocks $A_{t,j}$ and $A_{j,t}$ as well as the tip block $A_{t,t}$.
In Fig.~\ref{fig:schur_bta_i}(b), we provide a compressed block representation of the operations to be executed at step $i$ of the extended \textsc{RGF} algorithm, avoiding the explicit zero computations specific to the BTA sparsity pattern.
% The exact derivation of both algorithms is deferred to the Supplementary Material~\ref{}.
The forward Schur complement for the quadratic matrix equation involves $34$ gemm, $1$ getrf (LU), and $2$ trsm block-operations per step. 
The associated backward selected quadratic solve requires $76$ gemm block-operations per step.

\subsection{Parallel Algorithms for the Selected Solution of the Quadratic Equation}
We derive here parallel methods to solve $AXA^\dagger=B$ for BT and BTA matrices. 
It is assumed that $A$ and $B$ possess the same sparsity pattern. 
Note that BT matrices can be treated as a special case of BTA ones with $a$=$0$.

\subsubsection{Overview}
While the block approach of \textsc{RGF} is ideally suited to leverage intra-block parallelism through level-3 BLAS operations, the algorithm itself presents inherent sequential dependencies between subsequent block operations.
In order to break these sequential dependencies at the block level and reduce the depth of the algorithm, we introduce a permutation scheme that gives rise to embarrassingly parallel sections at the cost of more block operations and thus increased workload.
These embarrassingly parallel sections are distributed among $P$ processes (CPUs or GPUs).

Our scheme adds work in the form of fill-in induced by the permutation scheme and of a reduced system of equations to be solved that connects the embarrassingly parallel partitions.
We denote as ``middle partitions'' all $P_i$ connected upward (downward) to a partition $P_{i-1}$ ($P_{i+1}$). 
The first and last partitions, $P_0$ and $P_{P-1}$, are only connected to $P_1$ and $P_{P-2}$, respectively.
Since the connections are different for the middle and first/last partitions, so is the additional work. Hence, the distributed algorithm requires load balancing that can be achieved by adjusting the number of diagonal blocks in the middle partitions.

\subsubsection{Permutation}\label{sec:permutation}
\begin{figure}[t]
    \centering
    \includegraphics[width=0.95\columnwidth]{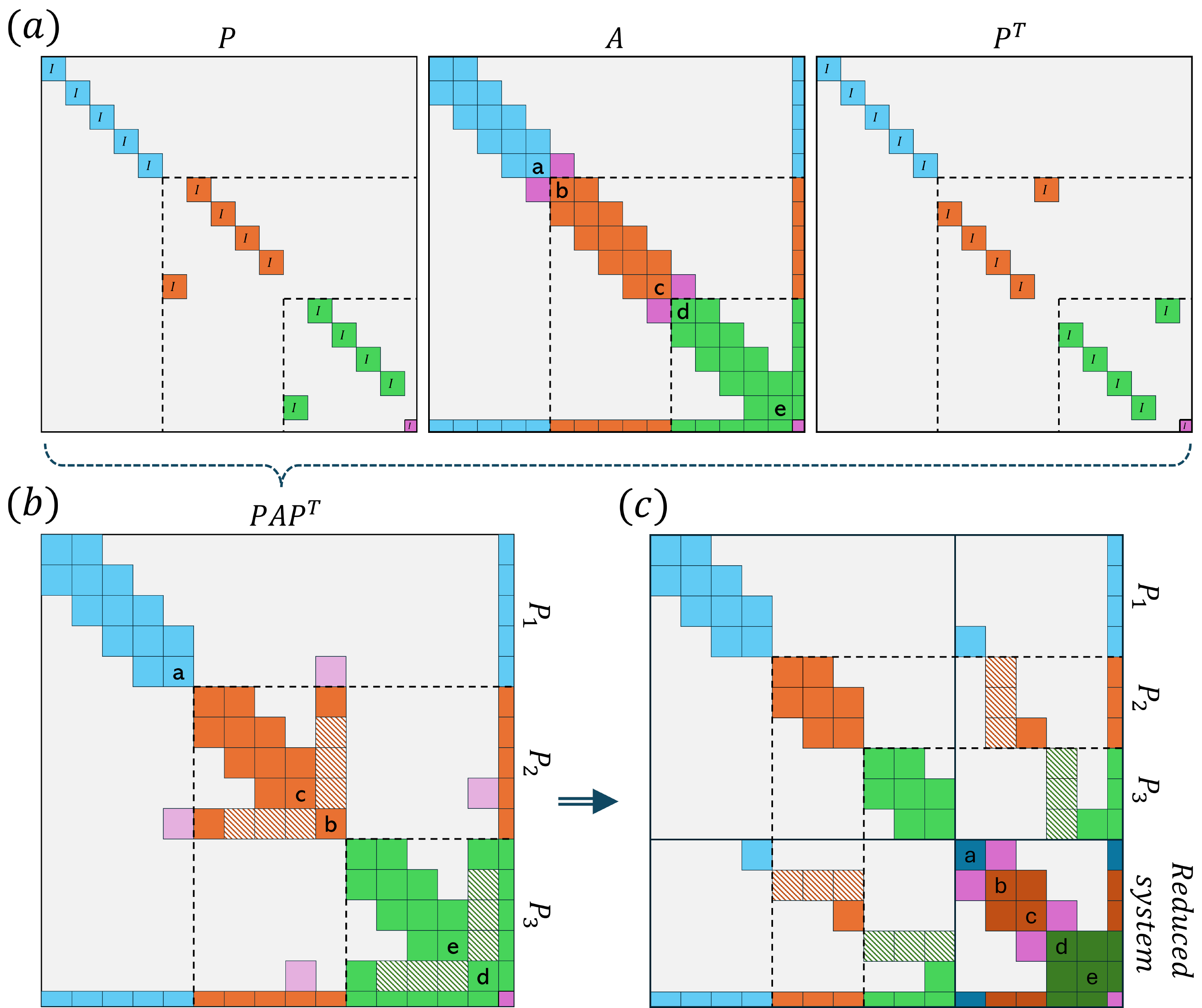}
    \caption{(a) Graphical representation of the permutation $PAP^T$ where the matrix $A$ is distributed over three processes. All partitions have a different color. (b) Result of the permutation and re-ordering of the partitioned matrix $A$ according to the operation in (a). The hatches refer to the fill-in induced by the permutation during the Schur complement operation.}
    \label{fig:permutation_distributed}
\end{figure}

The permutation of the $A$ and $B$ matrices introduced to break the block sequential dependencies of \textsc{RGF} is performed implicitly, without memory copy, but by reordering the block access pattern at the algorithmic level.
% ---
The permutation matrix $P$ in Fig.~\ref{fig:permutation_distributed}(a) is used for that purpose. In this specific case, it leads to a domain decomposition or dissection of the BTA matrix considered into
three partitions delimited by dashed lines and colored in \textcolor{a_blue}{blue}, \textcolor{b_orange}{orange}, and \textcolor{a_green}{green}.
The blocks connecting two partitions of $A$ serve as separators. They are represented in \textcolor{purple}{purple} color and couple, for example, the blocks \colorbox{a_blue}{\emph{a}} and \colorbox{b_orange}{\emph{b}} of $A$.

Figure~\ref{fig:permutation_distributed}(b) gives a visual representation of the reordered matrix $PAP^T$. % after permutation. 
The decomposition fill-in induced by the permutation in the middle partition is highlighted in hatches.
Finally, in Fig.~\ref{fig:permutation_distributed}(c), the permuted matrix is reordered to isolate the embarrassingly parallel sections that have been exposed (upper left quadrant \includegraphics[height=1.4ex]{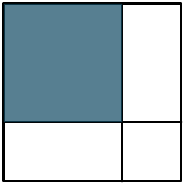}), the sequential reduced system (lower right quadrant \includegraphics[height=1.4ex]{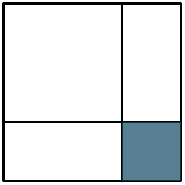}), and the decomposition fill-in (upper right \includegraphics[height=1.4ex]{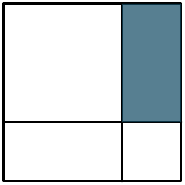} and lower left \includegraphics[height=1.4ex]{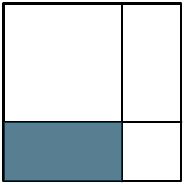} quadrants).
% ---
In the context of the quadratic matrix equation, the partitioning and implicit permutation are applied to both the system matrix $A$ and the right-hand-side matrix $B$.

\subsubsection{Intuitive Derivation}
% These 2 figures belong together ---
\begin{figure*}[t]
    \centering
    \caption{Quadratic matrix equation $AXA^\dagger=B$ after permutation and reordering. At step $i$, the Schur complement involves the sliced block multiplication in the light-orange, light-brown, and light-violet regions. The gray arrows indicate the forward updates preparing the complement operation of the next step.}
    \includegraphics[width=0.9\textwidth]{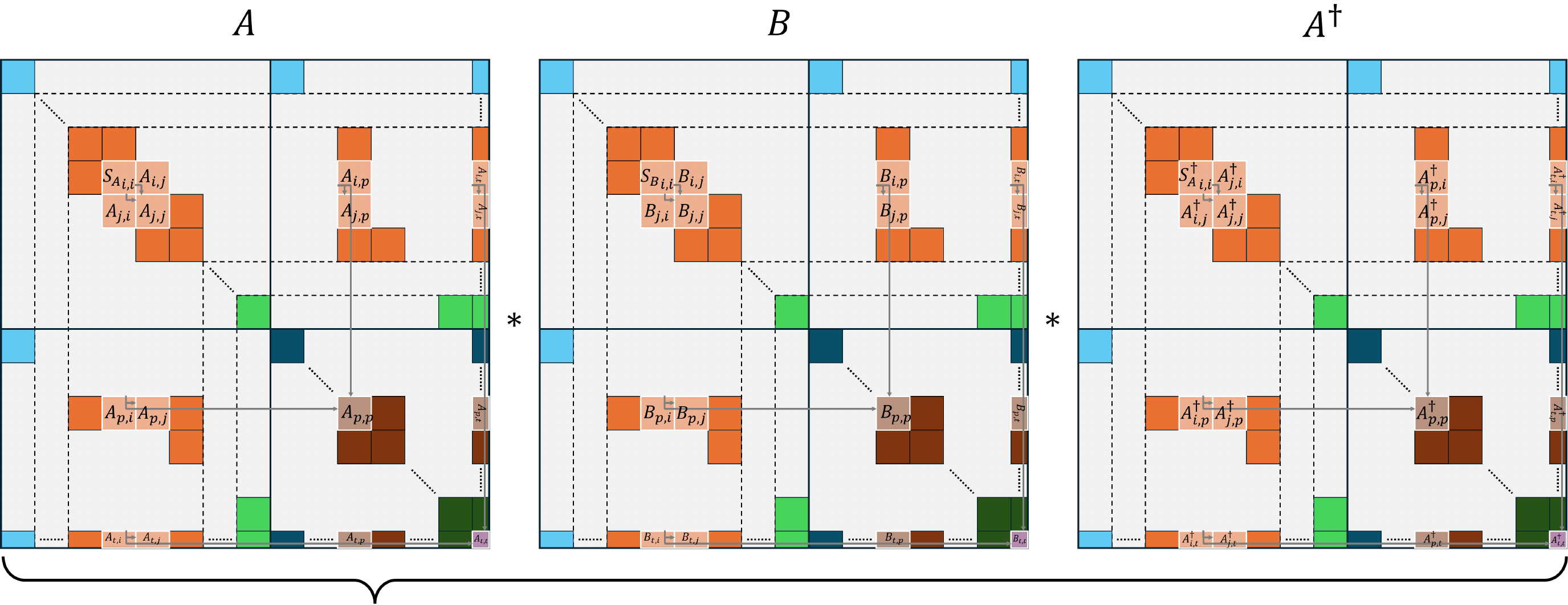}
    \label{fig:permutation_quadratic}
\end{figure*}
\begin{figure}[t]
    \centering
    \vspace{-25pt}
    \includegraphics[width=0.8\columnwidth]{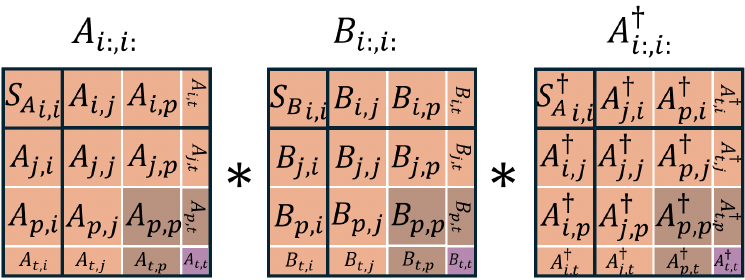}
    \caption{Compressed structure extracted from Fig.~\ref{fig:permutation_quadratic} highlighting the Schur complement operations and the forward updates that are computed at each iteration $i$.}
    \label{fig:quadratic_compressed}
\end{figure}
% -----------------------------------
Similarly to the BTA extension of the \textsc{RGF} algorithm, the full algorithm derivation includes many inter-dependent block operations.
We provide here a visual intuition on the derivation of the distributed selected inversion/solution algorithm introduced in this work.
% ---
Our approach allows one to compute the Schur complement and selected solution of $AXA^\dagger=B$ alongside the selected inversion of $AX=I$.
In Fig.~\ref{fig:permutation_quadratic}, the permutation presented in the previous section is applied to the Schur complement of the quadratic matrix equation.
On top of the permutation, the block operations that must be performed at a given step $i$ of the algorithm, for a middle partition, are highlighted.
These operations lead to the computation of the Schur complement of the matrix $B$: $S_{B_{j:,j:}} = S_{A_{i:,i:}} S_{A_{i:,i:}} S_{A_{i:,i:}}^{\dagger}$.

Following the slicing and reordering introduced in Section~\ref{sec:permutation}, we identify in Fig.~\ref{fig:permutation_quadratic} three types of operation: (i) update of the next diagonal block $B_{j,j}$, (ii) update of the matching diagonal block $B_{p,p}$ inside the reduced system, and (iii) update of the tip of the arrowhead $B_{t,t}$, also part of the reduced system.
This access pattern is the same for all processes except for the middle ones, as the blocks at the boundary of their partition, that is, those connecting them to their upper/lower neighbors, are also updated.
Following the same ideas as in Fig.~\ref{fig:schur_bta_i}, we compress the representation of the block operations by avoiding the \textit{a priori} zeros computation.
This compressed block representation is presented in Fig.~\ref{fig:quadratic_compressed}. 
The updated blocks highlighted in darker tones represent the local contribution of the current process to the reduced system.

Once the embarrassingly parallel Schur complement has been computed on all processes, the reduced system can be gathered locally.
It is then solved either by using the BTA \textsc{RGF} algorithm presented in Section~\ref{ssec:bta_schur} or by a recursive call to the distributed version of the algorithm.
This operation ultimately constitutes the sequential bottleneck of the algorithm.
Finally, the slicing and block dependencies are reverted, and the embarrassingly parallel selected inversion is performed.
Each process produces the blocks of the inverse/solution matching the non-zero pattern of its partition of the system matrix.
Overall, this constitutes the complete and distributed quadratic selected solution of the original linear system.

\subsection{Analysis}
We present in Table~\ref{tab:analysis_methods} the asymptotic computational complexity and memory footprint of the distributed-memory algorithms.
We separate the contribution from the \textcolor{Emerald}{embarrassingly parallel} sections and from the \textcolor{Bittersweet}{additional work} induced by the permutation. 
The computational efficiency of the distributed algorithms reaches its optimum when $\frac{n}{P} \gg P$, i.e., when the size of each partition is greater than the number of processes.

We present in Table~\ref{tab:op_count} the counts of the block operations per step (number of diagonal blocks) for all algorithms presented in this paper.
% The newly introduced numerical methods are colored in \textcolor{Bittersweet}{red}.
The extension of the \textsc{RGF} algorithm to BTA matrices comes at the cost of computing, for each step $i$, several additional matrix-matrix products of shapes $abb$, $aab$, $bba$, $baa$, $aba$, and $bab$.
In practice, for many scientific applications, the size of the arrowhead $a$ is small compared to the diagonal blocksize $b$, which leads to the computational expenses being dominated by the matrix-matrix product of shape $bbb$.

\begin{table}[t]
\centering
\resizebox{\columnwidth}{!}{%
\begin{tabular}{@{}lccccccccc@{}}
\toprule
\multicolumn{1}{c}{} & \multicolumn{7}{c}{\textbf{MM (mnk)}} & \textbf{LU} & \textbf{TRSM} \\ \cmidrule(l){2-10} 
\multicolumn{1}{c}{\multirow{-2}{*}{\textbf{Algs}}} & $abb$ & $aab$ & $bba$ & $baa$ & $aba$ & $bab$ & $bbb$ & $b$ & $b$ \\ \midrule
{\color[HTML]{000000} ddbt{\color[HTML]{FE0000}a}sc} & {\color[HTML]{FE0000} 1} & {\color[HTML]{FE0000} 0} & {\color[HTML]{FE0000} 2} & {\color[HTML]{FE0000} 0} & {\color[HTML]{FE0000} 1} & {\color[HTML]{FE0000} 0} & 2 & 1 & 2 \\
ddbt{\color[HTML]{FE0000}a}sc (q) & {\color[HTML]{FE0000} 5} & {\color[HTML]{FE0000} 0} & {\color[HTML]{FE0000} 5} & {\color[HTML]{FE0000} 0} & {\color[HTML]{FE0000} 4} & {\color[HTML]{FE0000} 0} & 8 & 1 & 2 \\
ddbt{\color[HTML]{FE0000}a}sc (perm) & {\color[HTML]{FE0000} 3} & {\color[HTML]{FE0000} 0} & {\color[HTML]{FE0000} 2} & {\color[HTML]{FE0000} 0} & {\color[HTML]{FE0000} 1} & {\color[HTML]{FE0000} 0} & 6 & 1 & 2 \\
ddbt{\color[HTML]{FE0000}a}sc (perm) (q) & {\color[HTML]{FE0000} 10} & {\color[HTML]{FE0000} 0} & {\color[HTML]{FE0000} 8} & {\color[HTML]{FE0000} 0} & {\color[HTML]{FE0000} 4} & {\color[HTML]{FE0000} 0} & 22 & 1 & 2 \\ \midrule
ddbt{\color[HTML]{FE0000}a}sci & {\color[HTML]{FE0000} 2} & {\color[HTML]{FE0000} 1} & {\color[HTML]{FE0000} 2} & {\color[HTML]{FE0000} 1} & {\color[HTML]{FE0000} 0} & {\color[HTML]{FE0000} 3} & 7 & 0 & 0 \\
ddbt{\color[HTML]{FE0000}a}sci (q) & {\color[HTML]{FE0000} 6} & {\color[HTML]{FE0000} 3} & {\color[HTML]{FE0000} 9} & {\color[HTML]{FE0000} 4} & {\color[HTML]{FE0000} 0} & {\color[HTML]{FE0000} 13} & 34 & 0 & 0 \\
ddbt{\color[HTML]{FE0000}a}sci (perm) & {\color[HTML]{FE0000} 3} & {\color[HTML]{FE0000} 1} & {\color[HTML]{FE0000} 5} & {\color[HTML]{FE0000} 1} & {\color[HTML]{FE0000} 0} & {\color[HTML]{FE0000} 5} & 14 & 0 & 0 \\
ddbt{\color[HTML]{FE0000}a}sci (perm) (q) & {\color[HTML]{FE0000} 10} & {\color[HTML]{FE0000} 3} & {\color[HTML]{FE0000} 14} & {\color[HTML]{FE0000} 5} & {\color[HTML]{FE0000} 0} & {\color[HTML]{FE0000} 22} & 72 & 0 & 0 \\ \bottomrule
\end{tabular}
}
\caption{Count of the block operations for the different algorithms presented in this work. The keyword \emph{perm} signifies that a permuted algorithm is used in the distributed versions, whereas the keyword \emph{q} refers to the selected solution of the quadratic matrix equation. The operations highlighted in {\color[HTML]{FE0000}red} are only needed for the BTA version of the algorithm.}
\label{tab:op_count}
\end{table}

\begin{table}[h]
\centering
\resizebox{\columnwidth}{!}{%
\begin{tabular}{@{}lcc@{}}
\toprule
\multicolumn{1}{c}{\multirow{2}{*}{\textbf{\begin{tabular}[c]{@{}c@{}}Sparsity\\ Pattern\end{tabular}}}} & \multicolumn{2}{c}{\textbf{Distributed}} \\ \cmidrule(l){2-3} 
\multicolumn{1}{c}{} & \textit{Compute} & \textit{Memory} \\ \midrule
BT & $O($\colorbox{Emerald!40}{$\frac{nb^3}{P}$}$+$\colorbox{Bittersweet!45}{$Pb^3$}$)$ & $O($\colorbox{Emerald!40}{$\frac{nb^2}{P}$}$+$\colorbox{Bittersweet!45}{$Pb^2$}$)$ \\
BTA & $O($\colorbox{Emerald!40}{$\frac{nb^3+nba^2+a^3}{P}$}$+$\colorbox{Bittersweet!45}{$P(b^3+ba^2+a^3)$}$)$ & $O($\colorbox{Emerald!40}{$\frac{nb^2+nba+a^2}{P}$}$+$\colorbox{Bittersweet!45}{$P(b^2+ba+a^2)$}$)$ \\ \bottomrule
\end{tabular}
}

\begin{tikzpicture}
    \fill[Emerald!40] (0,0) rectangle (0.25cm,0.25cm);
    \node[right] at (0.25cm,0.125cm) {Embarassingly parallel.};

    \fill[Bittersweet!45] (4cm,0) rectangle (4.25cm,0.25cm);
    \node[right] at (4.25cm,0.125cm) {Reduced system work.};
\end{tikzpicture}

\caption{Asymptotic complexity and memory footprint analysis of the proposed parallel algorithms.}
\label{tab:analysis_methods}

\end{table}

\begin{figure*}[t]
    \centering
    \includegraphics[width=1\textwidth]{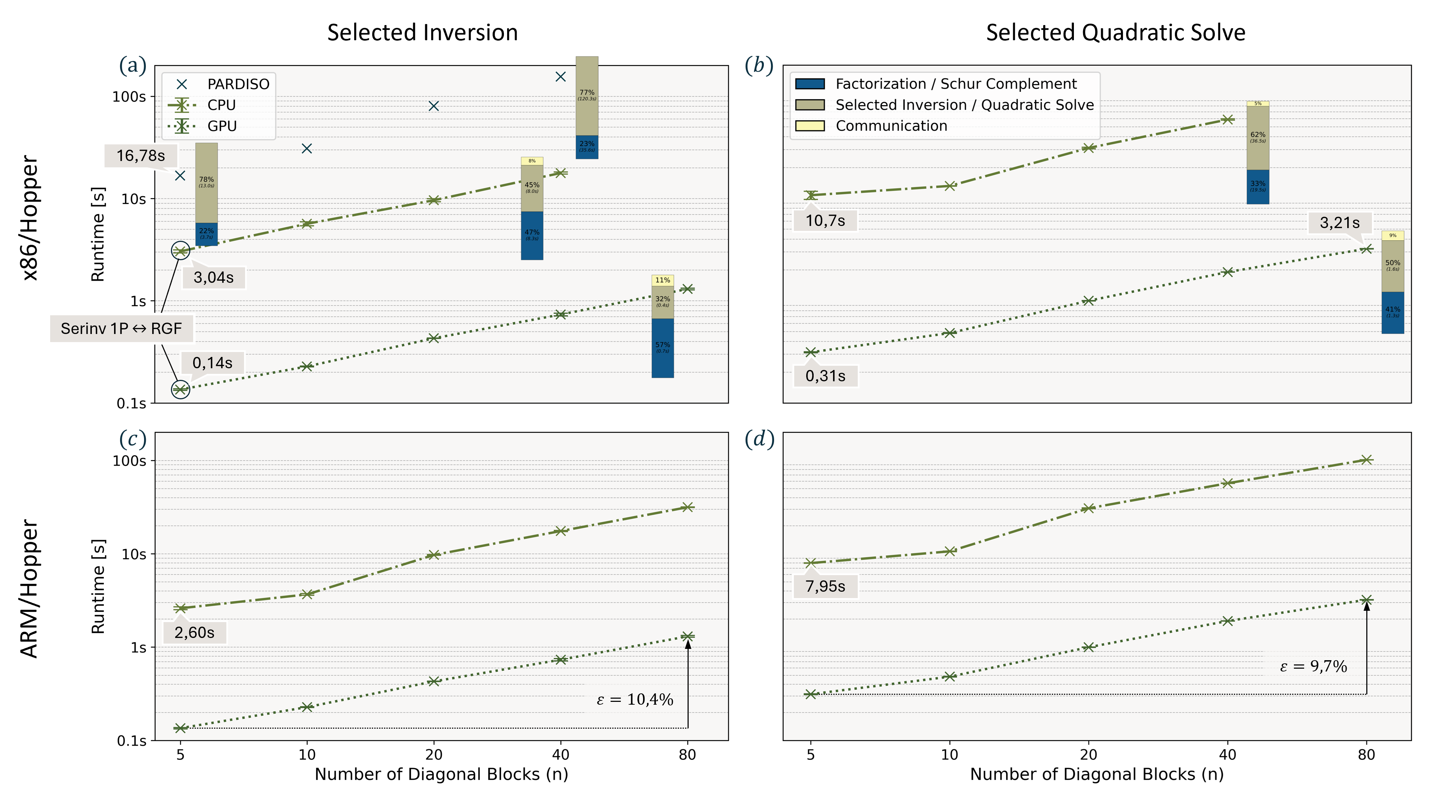}
    \caption{Weak-scaling performance of our approach on a real dataset from a QT simulation of an NR-FET transistor as a function of the number $n$ of diagonal blocks. The diagonal block size $b$ is set to 3408. (a) and (c) Selected inversion. (b) and (d) Fused selected inversion and selected solution of the quadratic problem. Comparisons (when possible) with \textsc{PARDISO} in its optimal multithreaded configuration ($OMP=64$) are provided. Evaluations were performed on (a-b) x86/Hopper and (c-d) ARM/Hopper architectures. Stacked bar plots show the percentages of runtimes spent in factorization, selected inversion / quadratic solve, and, when applicable, communication.}
    \label{fig:weak_Scaling_serinv}
\end{figure*}
\begin{figure*}[t]
    \centering
    \includegraphics[width=1\textwidth]{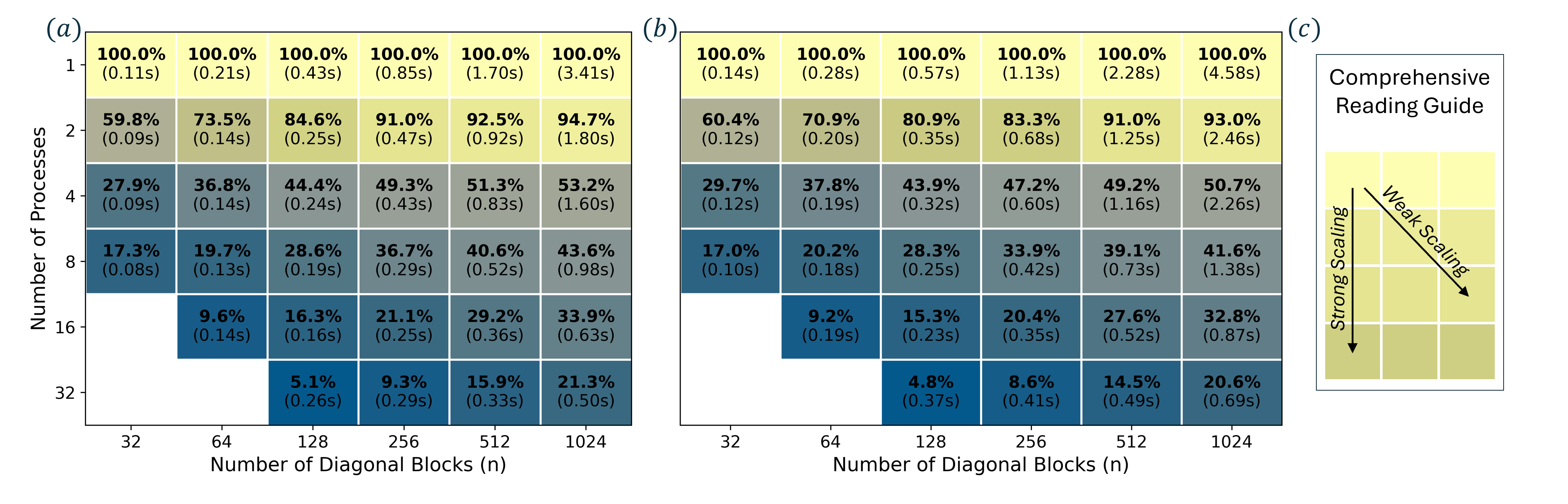}
    \caption{Evaluation of the parallel efficiency of our GPU implementations on GH200 superchips on the synthetic dataset $SD$. (a) Selected inversion. (b) Fused selected inversion and selected quadratic solve. The problem sizes include $n=[32, 64, 128, 256, 512, 1024]$) diagonal blocks, all with a size $b=1024$. (c) Reading guide to interpret (a) and (b).}
    \label{fig:perfmatrix}
\end{figure*}

\begin{table}[b]
\centering
\resizebox{\columnwidth}{!}{%
\begin{tabular}{@{}lccc@{}}
\toprule
\textbf{Reference} & \multicolumn{1}{c}{\textbf{n}} & \textbf{b} & \textbf{N} \\ \midrule
$NR_{3408}$ & $\{5,10,20,40,80\}$ & $3408$ & $17k - 272k$ \\
$SD$ & $\{32,64,128,256,512,1024\}$ & $1024$ & $32k - 1M$ \\ \bottomrule
\end{tabular}
}
\caption{Datasets used in the performance evaluations of our distributed algorithms. Here, $n$ is the number of main diagonal blocks, $b$ is the diagonal block size, and $N$ is the total matrix dimension.}
\label{tab:datasets}
\end{table}

The distributed algorithms rely on an aggregation of the updated blocks from each embarrassingly parallel partition into a reduced system that must be treated sequentially.
Each middle process sends to all its counterparts $2$ diagonal blocks of size $b^2$, $2$ upper/lower diagonal blocks of sizes $b^2$ as well as $4$ lower (upper) arrowhead blocks of size $ab$ with the \emph{AllGather} collective.
The tip of the reduced system, which encompasses the updated arrowhead block, is built from data originating from all processes and aggregated through the \emph{AllReduce} collective of size $a^2$.

\section{Evaluation}\label{sec:evaluation} % 6 Columns

\subsection{Methodology}
We benchmarked the performances of our approaches in two ways.
First, we used a real quantum transport dataset, labeled $NR_{3408}$, as benchmark. It corresponds to an NR-FET, similar to the one presented in Fig.~\ref{fig:hamiltonians}. The length of the device's channel was increased proportionally to the number of processes to test our distributed algorithms. 
Second, we determined the parallel efficiency of our approach based on a synthetic dataset, labeled $SD$, that covers a wide range of possible channel lengths.
A summary of the datasets is presented in Table~\ref{tab:datasets}.

\begin{table}[t]
\centering
\resizebox{\columnwidth}{!}{%
\begin{tabular}{@{}lccl@{}}
\toprule
\multicolumn{1}{c}{\textbf{Name}} & \textbf{Architecture} & \multicolumn{1}{c}{\textbf{Description}} & \multicolumn{1}{c}{\textbf{Memory}} \\ \midrule
FAU & \begin{tabular}[c]{@{}c@{}}x86\\ (Sapphire Rapids)\end{tabular} & \begin{tabular}[c]{@{}l@{}}2 x Intel Xeon Platinum 8470 \\ (2 x 52 cores @ 2.0 GHz)\end{tabular} & 2TB DDR5 \\
ALPS & \begin{tabular}[c]{@{}c@{}}ARM (Grace) \\ Hopper\end{tabular} & \begin{tabular}[c]{@{}l@{}}Nvidia GH200 \\ (72-core CPU + 1 GPU)\end{tabular} & \begin{tabular}[c]{@{}l@{}}128GB LPDDR5X \\ 96GB HBM3\end{tabular} \\ \bottomrule
\end{tabular}
}
\caption{Summary of the hardware used to evaluate the performance of our algorithm implementations.}
\label{tab:hardware}
\end{table}

In both investigations, we measured the time-to-solution of our selected inversion and selected quadratic solution methods, and compared the results, when possible, to PARDISO. The mean of at least 10 runs, along with the 95\% confidence interval are reported in all cases. Two different hardware architectures were employed: x86 CPUs combined with Hopper GPUs and ARM CPUs plus Hopper GPUs (GH200 superchips), as provided by the Erlangen National High Performance Computing Center (NHR@FAU) and the Swiss National Supercomputing Center (CSCS), respectively. Their features are summarized in Table~\ref{tab:hardware}. We used MPI (CPU) and NCCL (GPU) as communication libraries for the distributed codes. 

The CPU evaluations were performed with the most favorable multithreading settings: 32 OpenMP (OMP) threads on the x86 system and 64 OMP threads on ARM for our approach, 64 OMP threads for PARDISO on x86. 
For PARDISO, the input matrices were reordered using nested-dissection from the METIS package~\cite{metis_manual}, which was found to be the best-suited approach for the sparsity patterns considered. Additionally, a memory pool of up to 2~TB was allocated.

\subsection{Weak Scaling of the NR-FET Device}
Figure~\ref{fig:weak_Scaling_serinv} presents the weak scaling evaluation of our distributed algorithms for the $NR_{3408}$ dataset: selected inversion in sub-plots (a) and (c), fused selected inversion and selected quadratic solution in sub-plots (b) and (d). The time-to-solution results are reported for CPUs (x86 and ARM) and Hopper GPUs.
We also provide measurements for the selected inversion solver of \textsc{PARDISO} on x86 CPUs, the only configuration available for it, using the best-performing number of threads (64). Since this number is not adjusted with respect to the matrix size, the \textsc{PARDISO} timings in Fig.~\ref{fig:weak_Scaling_serinv} do not represent a weak-scaling experiment.

\subsubsection{Selected Inversion} For the shortest channel length of the $NR_{3408}$ device with $n=5$ blocks, \textsc{PARDISO} completes the selected inversion in $16.78\text{s}$, whereas our approach on one x86 (ARM) CPU takes $3.04$ ($2.6$) $\text{s}$, which corresponds to a $5.5\times$ ($6.45\times$)  speed-up on a single node. 
The GPU implementation performs the same computation in $0.14\text{s}$, i.e., $120\times$ faster than \textsc{PARDISO} and $22.5\times$ faster than its CPU counterpart on x86. 

When increasing the number of diagonal blocks to $n=40$ and the number of nodes to 8, our approach completes the selected inversion in $17.76\text{s}$ ($17.72\text{s}$) on the x86 (ARM) CPUs, about the same time that \textsc{PARDISO} needs to process the shortest device. Our implementation spends $47\%$ of the time on factorization, $45\%$ on inversion, and $8\%$ on communication. The selected inversion of the same device with \textsc{PARDISO} requires $156\text{s}$ on 1 node. 
Our GPU code computes the inversion of the largest device investigated ($n=80$) on 16 nodes in $1.3\text{s}$, spending $57\%$ of the time on factorization, $32\%$ on selected inversion, and $11\%$ on communication.
% -----------------------------------
When scaling from $n=5$ (single CPU or GPU) to $n=80$ (16 CPUs or GPUs), our approach on ARM CPUs and Hopper GPUs achieves a parallel efficiency of $\eta = 8.26\%$ and $\eta = 10.4\%$, respectively.

\subsubsection{Selected Quadratic Solution} For the shortest device ($n=5$), the selective quadratic solution fused with the selected inversion completes in $10.7\text{s}$ ($7.95\text{s}$) on one x86 (ARM) CPU and $0.31\text{s}$ on one GPU. Hence, such simulations, which allow to account for scattering, run $1.57$ to $2.11\times$ (CPU) and $54.1\times$ (GPU) faster than \textsc{PARDISO} and its selected inversion solver. 
At $n=80$ ($16$ GPUs), the selected quadratic solution with our approach completes in $3.21\text{s}$, which is $5.22\times$ faster than \textsc{PARDISO}'s selected inversion of a device $16\times$ shorter. 

\subsection{Parallel Efficiency Matrices}
We present in Fig.~\ref{fig:perfmatrix} parallel efficiency matrices of the GPU implementation of our methods on the NVIDIA GH200 superchips of the ALPS supercomputer for (a) the selected inversion and (b) fused selected inversion and selected solution of the quadratic matrix equation on dataset $SD$.
Keeping the size of the diagonal blocks constant ($b=1024$), we scale the number of diagonal blocks (horizontal axis) and the number of GPUs (vertical axis). A guide to interpret the performance matrix is provided in sub-plot (c).

The weak scaling of the selected inversion from $n=32$ on a single GPU ($0.11s$) to $n=1024$ on 32 GPUs ($0.5s$) exhibits a parallel efficiency $\eta=21.3\%$. The same experiment for the fused selected inversion and selected solution of the quadratic problem reaches a parallel efficiency $\eta=20.6\%$.
Such a performance showcase the successful extension of the existing distributed methods for the selected inversion of BT matrices to the selected solution of quadratic matrix equation.

\section{Discussion}\label{sec:conclusion} % 1 Column
We introduced distributed-memory algorithms for the selected inversion and selected solution of the quadratic matrix equation arising from quantum transport simulations of nano-devices based on the non-equilibrium Green's function (NEGF) formalism. 
Building upon Ref.~\cite{petersen_hybrid_2009} and extending the well-established recursive Green’s function algorithm, our methods can deal with both block-tridiagonal and block-tridiagonal-with-arrowhead matrices.

Our implementations efficiently leverage distributed memory parallelism across CPU and GPU architectures. 
On real NR-FET transistor datasets, they achieve up to $120\times$ speed-up over the state-of-the-art sparse direct solver \textsc{PARDISO} for selected inversion. Moreover, they allow for the inclusion of scattering effects, which require the selected solution of a quadratic problem, while still being $5.2\times$ faster than the selected inversion solver of \textsc{PARDISO} applied to a device $16\times$ smaller. 
Weak- and strong-scaling analyses up to 32 GPUs demonstrate parallel efficiencies above 20\%, confirming that our algorithms can be used in practical device simulations where time-to-solution and memory consumption are critical factors. Hence, the proposed methods substantially broaden the applicability of NEGF-based quantum transport investigations to device sizes beyond the reach of traditional solvers.

\section*{Acknowledgments}
This work was supported by the Swiss National Science Foundation (SNSF) under grant $\mathrm{n^\circ}$ 209358 (QuaTrEx), grant $\mathrm{n^\circ}$ 200021 (NumESC), and by the Platform for Advanced Scientific Computing in Switzerland (BoostQT). We acknowledge the scientific support and HPC resources from CSCS under projects sm96 and lp16 as well as the Erlangen National High Performance Computing Center (NHR@FAU) of the Friedrich-Alexander-Universität Erlangen-Nürnberg (FAU) under the NHR project 80227.

\bibliographystyle{IEEEtran}
\bibliography{biblio}

\end{document}